\documentclass[
aps,%
12pt,%
final,%
notitlepage,%
oneside,%
onecolumn,%
nobibnotes,%
nofootinbib,%
superscriptaddress,%
noshowpacs,%
centertags]%
{revtex4}

\begin{document}
\selectlanguage{english}

\title{Detection of Giant Pulses from the Pulsar PSR B1112+50}

\author{\firstname{A.A.}~\surname{Ershov}}
\email{ershov@prao.psn.ru}
\affiliation{
Pushchino Radio Astronomy Observatory, Astrospace Center, Lebedev Physical Institute, 
}
\author{\firstname{A.D.}~\surname{Kuzmin}}
\email{akuzmin@prao.psn.ru}
\affiliation{
Pushchino Radio Astronomy Observatory, Astrospace Center, Lebedev Physical Institute, 
}


\begin{abstract}
We detected giant pulses from the pulsar PSR B1112+50. A pulse with an intensity that is a factor of 30 or more higher than the intensity of the average pulse is encountered approximately once in 150 observed pulses. The peak .ux density of the strongest pulse is about 180 Jy. This value is a factor of 80 higher than the peak .ux density of the average pulse. The giant pulses are narrower than the average prfile approximately by a factor of 5 and they cluster about the center of the average prfile. 

Key words: pulsars and black holes; neutron stars, giant pulses.
\end{abstract}

\maketitle

\section{Введение}
Giant pulses (GPs) are short-duration burst-like increases in the intensity of individual radio pulses from pulsars. These are rare events observed only in three pulsars: the Crab pulsar PSR B0531+21 (Staelin and Sutton 1970), the millisecond pulsars PSR B1937+21 (Wolszczan et al. 1984), and PSR B1821.24 (Romani and Johnston 2001). 

For normal pulsars, the intensity of individual pulses exceeds the intensity of the average pulse by no more than several factors (Hesse and Wielebinski 1974; Ritchings 1976). The GPs in the Crab pulsar PSR B0531+21 exceed the average level by a factor of 20 to $2\times 10^3$ (Lundgren et al. 1995). The GPs in the millisecond pulsar PSR B1937+21 and PSR B1821-24 exceed the average level by factor of 20 to 300 (Wolszczan et al. 1984; Cognard et al. 1996) and 20 to 80 (Romani and Johnston 2001), respectively. 

Kuzmin and Losovskii (2002) detected an extremely high brightness temperature $T_B\ge 4\times 10^{35}$ K of GPs in the millisecond pulsar PSR B1937+21. 

The GP duration is much shorter than the duration of the average pulses from this pulsar. The GP width for the millisecond pulsar PSR B1937+21 at frequencies of 430, 1420, and 2380 MHz does not exceed 10 mcs, which is a factors of 10 smaller than the width of the average profile, 100 mcs (Kinkhabwala and Thorsett 2000). The GP position is stable inside the average profile. 

A characteristic feature of the pulsars with GPs is a two-component intensity distribution: lognormal for most of the pulses and a power-law $N\propto S^\alpha$ for GPs with intensities above a certain level (Gower 1972; Lundgren et al. 1995). 

The power-law index for the Crab pulsar is $\alpha = -3.3$  (Argyle and Gower 1972; Lundgren et al. 1995). The boundary at which the distribution changes its pattern corresponds to about a 30-fold intensity of the average pulse. GPs account for about 2.5

Johnston et al (2001), Cramer et al. (2002), and Johnston and Romani (2002) reported the discovery of a new type of GP - microstructure in the Vela pulsar and PSR B1706.44. The intensity distribution of this microstructure also contains a power-law branch with indices of $-2.85$ and $-2.7$, respectively. However, in contrast to classical GPs (in PSR B1821.24, and PSR B1937+21), the excess intensity of the average pulse is less than a factor of 10 and 4 for the Vela pulsar and PSR B1706.44, respectively.

Attempts to find GPs in other pulsars (Phinney and Taylor 1979; Johnston and Romani 2002) have failed thus far.

We detected GPs from PSR B1112+50, which exhibit the characteristic features of GPs. 
\section{OBSERVATIONS AND DATA REDUCTION}
Our observations were carried out from January 9 through June 30, 2002, with the Large-Aperture Synthesis (BSA) Radio Telescope at the Pushchino Radio Astronomical Observatory of the Astrospace Center of the Lebedev Physical Institute. The telescope has an effective area of about 20000 square meters at zenith. One linear polarization was received. We used a 128-channel receiver with the channel bandwidth $\Delta f = 20$ kHz. The frequency of the first (highest frequency) channel was 111.870 MHz, the sampling interval was 0.922 ms, and the time constant was $\tau = 1$ ms. We observed individual pulses. The duration of one observing session was about 5 min; 180 pulses were observed during one session. 

We measured the GP flux densities using the calibration method based on noise measured from discrete sources with known flux densities in units of radio telescope flux sensitivity $\Delta S$. The pulsar peak flux density of the observed pulse $S_{max-obs}$ was determined from the relation 
\[S_{max~obs}=\delta S/k_{PSR}\times (\Delta f\times \tau)^{-1/2}\times (S/N),\]
where $k_{PSR}=sin(h_{PSR})$ is the factor that takes into account the dependence of the effective area of the radio telescope on pulsar height $h_{PSR}$, $\Delta f = 2.56$ MHz is the total system bandwidth, $\tau = 1$ ms is the time constant of the output device, and S/N is the signal-to-noise ratio. According to Kuzmin and Losovsky (2000), $\Delta S$ was represented as
\[\delta S=\delta S_{1000}\times (a+b\times T_{BB}),\]
where $\delta S_{1000}$ is the flux sensitivity of the radio telescope toward a sky region with the brightness temperature $T_{0}$ = 1000 . and $T_{bb}$ is the brightness temperature of the sky background toward the pulsar. We assumed $\Delta S_{1000}$ to be 100 mJy/MHz/s (Kutuzov 2000) and took a = 0.4 and b = 0.0006 from Kuzmin and Losovsky (2000). 
 
Over the above period, we carried out a total of 105 observing sessions containing 18900 pulsar periods. The peak flux density of the average pulse was $S_{max-obs}\simeq 2.3$ Jy. The period-averaged .ux density was $S\simeq 56$ mJy. 

We detected 126 pulses whose peak fluxes exceeded that of the pulse averaged over 105 days of observations by more than a factor of 30; for 17 of them (one pulse per 1100 observed pulses), this excess was more than a factor of 50. The total flux densities of 114 pulses exceeded the total flux density of the average pulse by a factor of 10 or more. 

Figure 1 shows two GPs as examples. Figure 1a shows pulse no. 31 of April 23 with a peak flux density of about 120 Jy, which is a factor of 50 higher than the peak flux density of the average pulse. 

Of the 126 pulses whose intensity exceeds the intensity of the average pulse by a factor of 30 or more, 36 belong to neighboring pairs or triples; hence, we can estimate the duration of enhanced pulsar activity to be several seconds. An example of such a GP group is shown in the lower figure, which presents the strongest individual pulse in all 105 days of observations (pulse no. 32 of January 16). This pulse has a peak flux density of about 180 Jy, which is a factor of 80 higher than that of the average pulse, and the total flux density of this pulse exceeds that of the average pulse by a factor of 16. The large increase in pulsar activity lasted for six periods, i.e., about 10 s. 

Figure 2 shows one of the GPs (January 16, 2002) in comparison with the profile averaged over 105 days of observations. The peak flux density of the average pulse is about 1/10 of the width of the noise track of the GP profiles. Therefore, we increased the average profile by a factor of 80 for convenience. TheGP width $w_{50} = 4$ ms $= 0.9^\circ$ at a 0.5 level and $w_{10} = 7$ ms $= 1.5^\circ$ at a 0.1 level. The mean width of the 126 GPs $w_{50} = 5.1 \pm 1.2$ ms $= 1.1 \pm 0.3^\circ$ at a 0.5 level and $w_{10} = 9 \pm 2$ ms $=2 \pm 0.4^\circ$ at 0.1 level.The dispersion broadening $\Delta t_{DM} = 1.1$ ms and the receiver constant 1 ms have virtually no effect on the GP width. The width of the pulsar profile averaged over 105 observing sessions is $w_{50} =24$ ms $=5.2^\circ$ and $w_{10} = 44$ ms $= 9.6^\circ$ at 0.5 and 0.1 levels, respectively. Thus, the GPs are narrower than the average pulse approximately by a factor of 5. 

Figure 3 shows the phases of the observed GPs relative to the center of the average profile. The positions of GPs are stable inside the average profile and they cluster in the middle part of the average profile. The phase difference between the GPs and the average profile is $\Phi_{GP} - \Phi_{AP} = -1 \pm 4$ ms $\simeq (-0.02 \pm 0.1 )\times w_{10}$. 
\[T_B=S\lambda ^2/2k\Omega,\]
where $\lambda$ is the wavelength of the received radio emission, $k$ is the Boltzmann constant, and $\Omega$ is the solid angle of the radio emitting region. Assuming the size of the radio emitting region to be $l\le c\times w_{50}$, where $c$ is the speed of light, and the distance to the pulsar to be $d = 0.54$ kpc (Taylor et al. 1995), we obtain $T_B\ge 10^{26}$ K for $w_{50} = 4$ ms and $S = 180$ Jy.

The GPs that we detected are not scintillations. At the frequency of our observations (111 MHz), the interstellar scintillation time scale (the time correlation radius) for a pulsar with the dispersion measure $DM = 9.16$ pc $cm^{-3}$ is about 1 min and 1000 days, respectively, for refraction and diffraction scintillations (Shishov et al. 1995), which significantly exceeds the duration of the observed GPs (less than several seconds). No GPs were detected in several pulsars with similar or lower dispersion measures observed during this period. 

\section{DISCUSSION}

The GPs that we detected from PSR B1112+50 exhibit the characteristic features of classical GPs from PSR B0531+21 and PSR B1937+21. The GP peak intensity exceeds the peak intensity of the average pulse by more than a factor of 30. The fraction of GPs (0.7

Figure 4 shows distribution of the ratio $R$ of the peak flux densities of individual pulses from PSR B1112+50 to the peak flux density of the average pulse (in equal bins of the logarithm of $R$) inferred from our observations. The distribution was constructed from 3320 pulses whose intensity exceeds the sensitivity threshold of our radio telescope. It exhibits the features of a two-component structure characteristic of GPs: a lognormal distribution for $I_{GP} / I_{AP} < 30$ and a power-law distribution for $I_{GP} / I_{AP} > 30$. The dashed line represents the least-squares-.tted lognormal distribution $F($lg$R) = A\times exp(-B\times ($lg$R-C)^2)$, where $A = 0.03$, $B = 9.4$, and $C = 1.03$. Here, $F$ is the ratio of the number of pulses with a given $log R$ to the total number of pulses. The solid line represents the power-law distribution $F\propto (I_{GP} / I_{AP})^\alpha$. The index $\alpha \simeq -3.6$ of the power-law part of the distribution needs to be refined because of poor statistics.

The absence in our observations of GPs exceeding the average pulse by more than a factor of 80 for PSR B0531+21 and PSR B1937+21 may stem from the relatively small (because of the long period of PSR B1112+50) number of observed periods, 18900 (compared to more than $10^6$ for PSR B0531+21 and PSR B1937+21). For these pulsars, pulses that exceed the average level by a factor of 80 occur approximately once in $10^5$ pulses. 

Note that PSR B1112+50 does not belong to the group of pulsars with the strongest magnetic .elds on the light cylinder, an integrated radio luminosity, and a potential di.erence in the polar-cap gap (Kuzmin 2002). Therefore, the GP of this pulsar may be different in nature. 

\section{CONCLUSIONS }

We detected GPs from the pulsarPSRB1112+50. At 111 MHz, the peak flux density of the strongest pulse is 180 Jy, which is almost a factor of 80 higher than the peak flux of the average pulse. Pulses whose peak intensities exceed the peak intensity of the average pulse by more than a factor of 30 occur approximately once in 150 periods. The GPs are approximately a factor of 5 narrower than the average profile and they cluster in the middle part of the average pro.le. The brightness temperature of the observed GPs is $T_B\ge 10^{26}$ K. 

\begin{acknowledgments}

We are grateful to V. V. Ivanova and A. S. Aleksandrov for the assistance during observations. This work was supported in part by the Russian Foundation for Basic Research (project nos. 01-02-16326 and 00-02-17447) and the Program of the Presidium of the Russian Academy of Sciences - Nonstationary Processes in Astronomy.

\end{acknowledgments}
\newpage
%
%

%
\newpage
\section{FIGURES CAPTIONS}

Fig. 1. Examples of GPs from PSR B1112+50: (a) - a single GP with a peak intensity of 120 Jy, (b) - the strongest GP with a peak intensity of about 180 Jy, which belongs to the series of six strong pulses. 

Fig. 2. The strongest GP from PSR B1112+50 shown with the average pulse. The amplitude of the average profile was increased by a factor of 80. 

Fig. 3. The phases of the observed GPs relative to the center of the average profile. 

Fig. 4. The distribution of peak intensities of individual pulses from PSR B1112+50. The ratio $R$ of the GP peak flux density to the peak flux density of the average profile is plotted along the $x$ axis. Equal intervals correspond to an increase in the GP flux density by the same factor. The ratio of the number of GPs exceeding the peak flux density of the average profile to the total number of observed pulses is plotted along the $y$ axis. The dashed line represents the least-squares-fitted lognormal distribution $F($lg$R) = A\times exp(-B\times ($lg$R-C)^2)$ and the solid line represents the power-law distribution $F\propto (I_{GP} / I_{AP})^\alpha$.

\newpage
\begin{figure}
\setcaptionmargin{5mm}
\includegraphics{fig1.ps} 
\end{figure}
\newpage
\begin{figure}
\setcaptionmargin{5mm}
\includegraphics{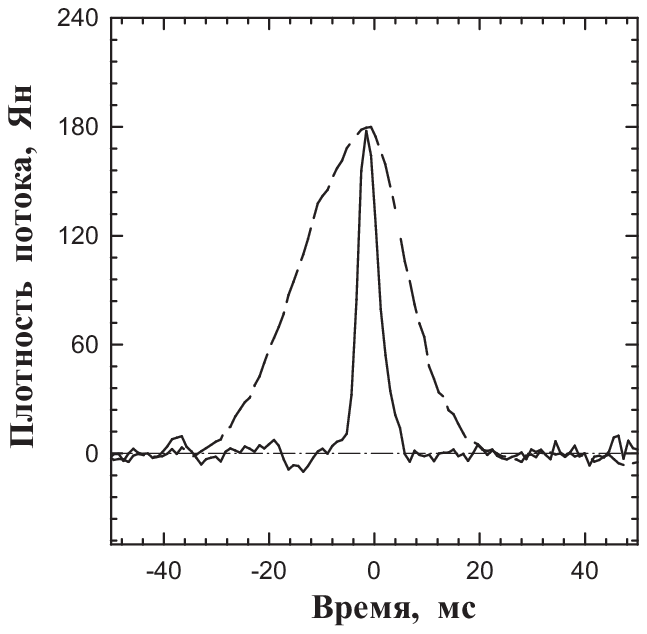} 
\end{figure}
\newpage
\begin{figure}
\setcaptionmargin{5mm}
\includegraphics{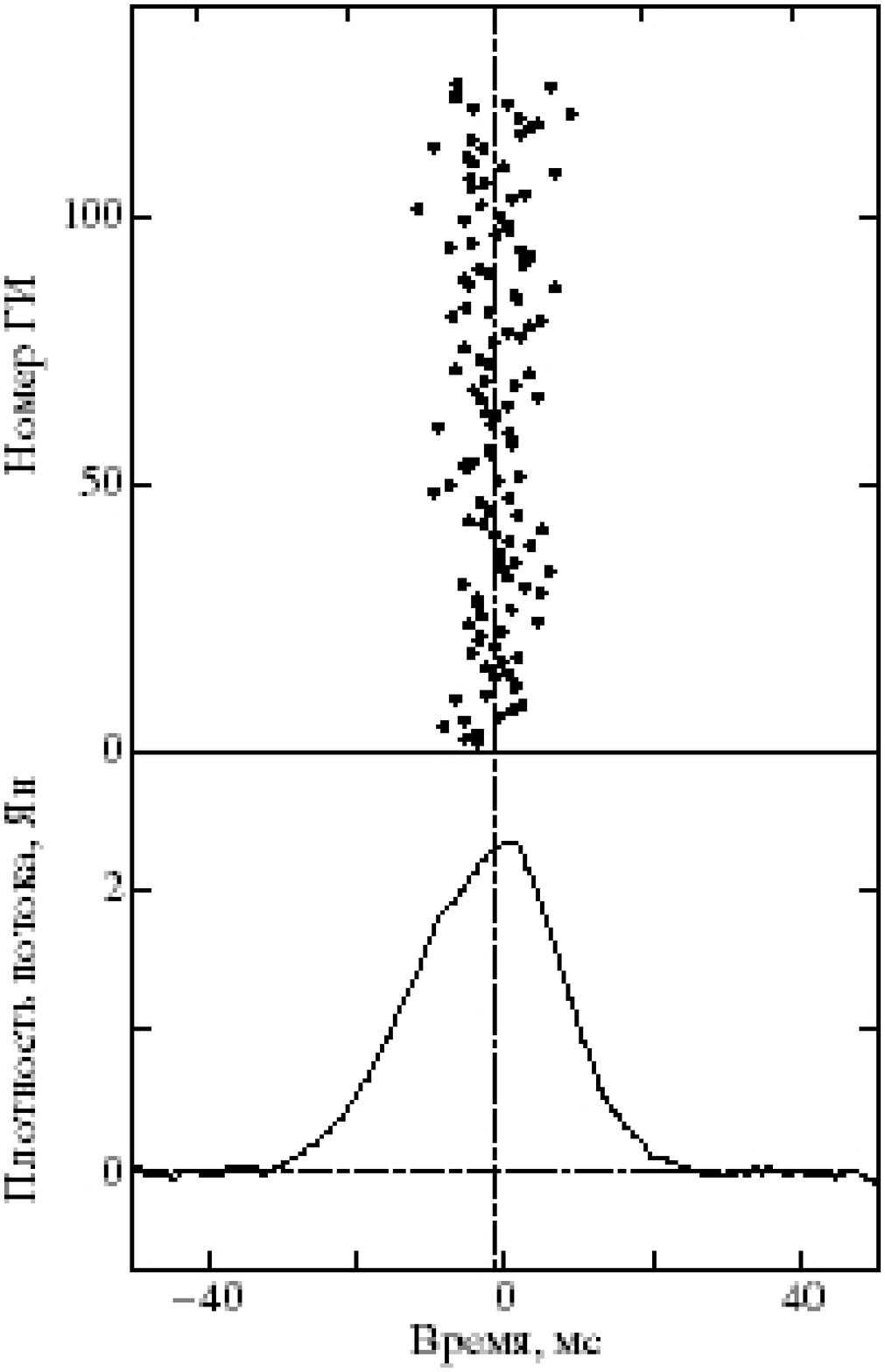} 
\end{figure}
\newpage
\begin{figure}
\setcaptionmargin{5mm}
\includegraphics{fig4.ps} 
\end{figure}
\end{document}